\documentclass[preprint]{aastex}
\begin{document}

\title{Detection of Radio Emission from Fireballs}
\author{K.S. Obenberger \\ \affil{Department of Physics and Astronomy, University of New Mexico, Albuquerque NM, 87131} G.B. Taylor\\  \affil{Department of Physics and Astronomy, University of New Mexico, Albuquerque NM, 87131}  J.M. Hartman\\ \affil{NASA Jet Propulsion Laboratory, Pasadena, CA 91109 USA \\ NASA Postdoctoral Program Fellow}  J. Dowell\\  \affil{Department of Physics and Astronomy, University of New Mexico, Albuquerque NM, 87131}  S.W. Ellingson\\ \affil{Bradley Dept. of Electrical Engineering, Virginia Tech, Blacksburg, VA, 24061} J.F. Helmboldt\\  \affil{US Naval Research Laboratory, Code 7213, Washington, DC 20375} P.A. Henning\\  \affil{Department of Physics and Astronomy, University of New Mexico, Albuquerque NM, 87131} M. Kavic\\  \affil{Department of Physics, Long Island University, Brooklyn, NY, 11201} F.K. Schinzel\\  \affil{Department of Physics and Astronomy, University of New Mexico, Albuquerque NM, 87131} J.H. Simonetti\\  \affil{Department of Physics, Virginia Tech, Blacksburg, VA, 24061} K. Stovall\\  \affil{Department of Physics and Astronomy, University of New Mexico, Albuquerque NM, 87131} T.L. Wilson\\  \affil{US Naval Research Laboratory, Code 7213, Washington, DC 20375}}

\begin{abstract} We present the findings from the Prototype All-Sky Imager (PASI), a backend correlator of the first station of the Long Wavelength Array (LWA1), which has recorded over 11,000 hours of all-sky images at frequencies between 25 and 75 MHz. In a search of this data for radio transients, we have found 49 long (10s of seconds) duration transients. Ten of these transients correlate both spatially and temporally with large meteors (fireballs), and their signatures suggest that fireballs emit a previously undiscovered low frequency, non-thermal pulse. This emission provides a new probe into the physics of meteors and identifies a new form of naturally occurring radio transient foreground.\\
\end{abstract}

\section{Introduction}
In recent years the field of radio transients has seen much interest, most of which has been focused on $>$ 1 GHz radio emission \citep{Frail12,Lorimer07,Keane12,Thornton13,Wayth12}. Only a handful of blind searches have been carried out below 100 MHz \citep{Cutchin11,Lazio10,Kardashev77} leaving this a relatively unexplored region of the spectrum.

A recent study with the LWA1 yielded two promising transients below 40 MHz \citep{Obenberger14}. This search was focused on placing limits on prompt emission from gamma ray bursts and therefore only the times shortly after the occurrence of GRBs were searched. In total $<$ 200 hours of data were analyzed. The two detected transients were not associated with GRBs, but at the time of publication their origins were unknown. However the need to conduct further investigation was evident. 

In this letter we present an analysis of over 11,000 hours of all-sky images recorded by the LWA1. Including the two previously mentioned events, a total of 49 transients have been detected in this data and there is strong correlation to large meteors known as fireballs. 

\section{Observations}
The LWA1 is a radio telescope located in central New Mexico, operating between 10 and 88 MHz. It consists of 256 cross-dipole antennas spread over an ellipse of 100 $\times$ 110 m. There are also 5 additional antennas at distances ranging from 200 to 500 m from the center of the array \citep{Ellingson13,Taylor12}. The telescope is collocated with the Karl Jansky Very Large Array (VLA), at a latitude of 34.070$^{\circ}$ N and a longitude of 107.628$^{\circ}$ W. 

PASI, a backend to the LWA1, correlates the signals from all antennas in real time, integrating for 5 s with 75 kHz bandwidth tunable to a center frequency anywhere within the 78 MHz over which the LWA1 operates. PASI also produces dirty images of the entire $\sim 2\pi$ sr sky above the LWA1 \citep{Obenberger14}, and in April 2012 began saving the images to a permanent archive\footnote{Prior to this the images were deleted after the generation of a movie comprising an entire day of images.}. This archive now contains 11,000 hours of all-sky images at various center frequencies. The vast majority of the data has been recorded at center frequencies of 37.8, 37.9, 52.0, and 74.0 MHz, with 30\%, 31\%, 17\%, and 13\% recorded at those frequencies, respectively.

\section{Discovery of Transients and Correlations with Fireballs}

Image subtraction algorithms have been developed to search for transients in the full 11,000 hours of PASI data. The underlying method is that from every image, the image 15 seconds prior is subtracted. Pixels above 6 $\sigma$ of the image noise are reported as candidate events and the equatorial coordinates are calculated. All four Stokes parameters and total spectra are saved and used to check if the candidate event is radio frequency interference (RFI). The coordinates of events are also checked against known bright sources which can be focused by the ionosphere at low frequencies \citep{Obenberger14}. 

This method has resulted in the discovery of 22 transients at 37.8 MHz, 20 at 37.9 MHz, 1 at 29.9 MHz, 1 at 25.6 MHz, and none at either 52.0 MHz or 74.0 MHz. The majority of these transients display a fast rise and exponential decay, lasting for tens of seconds to up to a few minutes, and have flux densities ranging from 500 to 3500 Jy (Fig. 1). They also show constant power across the 75 kHz band and contain low polarization levels consistent with instrumental leakage of unpolarized sources. 

\begin{figure}
	\centering
	\includegraphics[width = 7in]{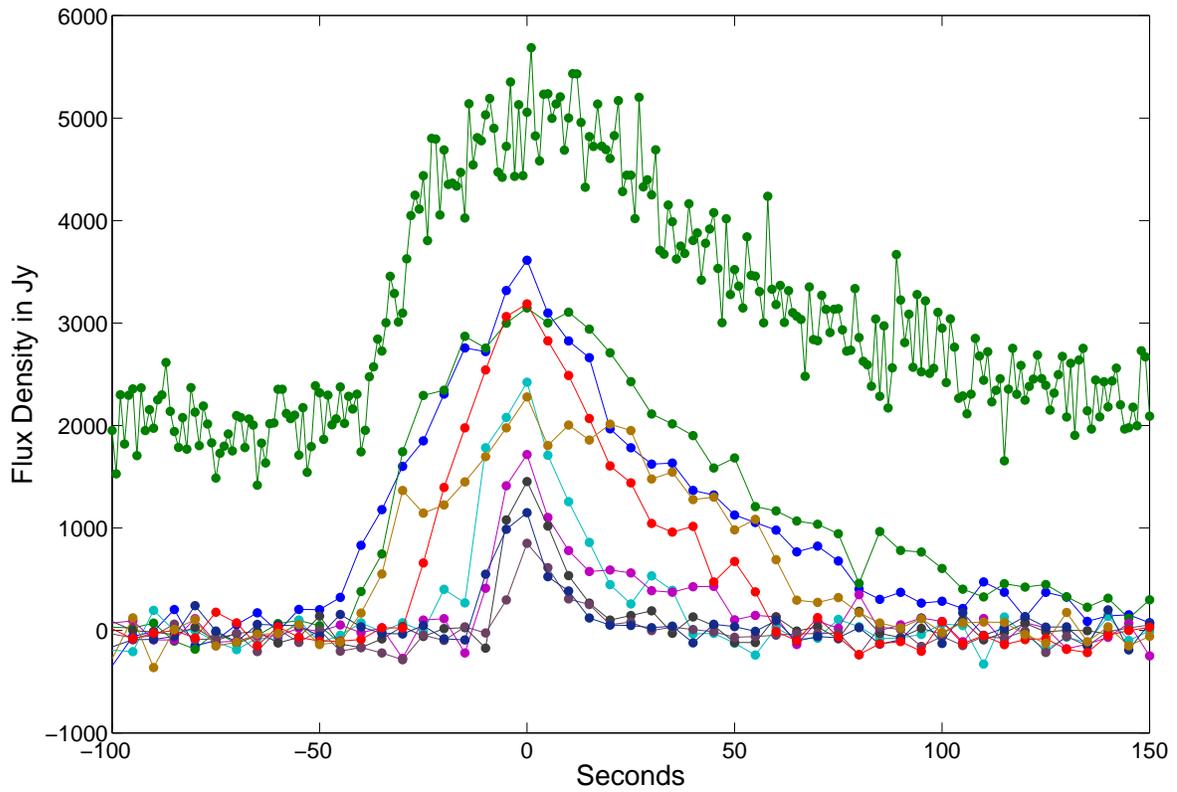}
	\caption{Light curves of the 9 brightest transients at 5 s integrations (Bottom). One of the transients shown on bottom (green) was also recorded using 1 s integrations (top), we show this light curve, offset by 2000 Jy, to illustrate that they this transient was smooth over 1 s timescales.}
\end{figure}

Most of the transients appear as point sources, meaning they are limited to $\leq4.4^{\circ}$ at 38 MHz. However some are extended over several degrees across the sky, most of which span less than ten degrees. In one case on January 21 2014,  a source leaves a trail covering 92$^{\circ}$ in less than ten seconds (Fig. 2). The trail then slowly recedes to the end point which glows for $\sim$90 seconds. The only known source that could cover this distance across the sky in less than 10 seconds and leave a persistent trail is a fireball. To investigate this further, the transients were compared with the detections from NASA's All Sky Fireball Network\footnote{http://fireballs.ndc.nasa.gov}. The network consists of 12 all-sky cameras, two of which are situated in Southern New Mexico and share a portion of the sky with the LWA1. The cameras are used to determine the 3 dimensional position, speed, absolute magnitude\footnote{Stellar magnitude at zenith}, and mass of the of the fireballs. 

\begin{figure}
	\centering
	\includegraphics[width = 7in]{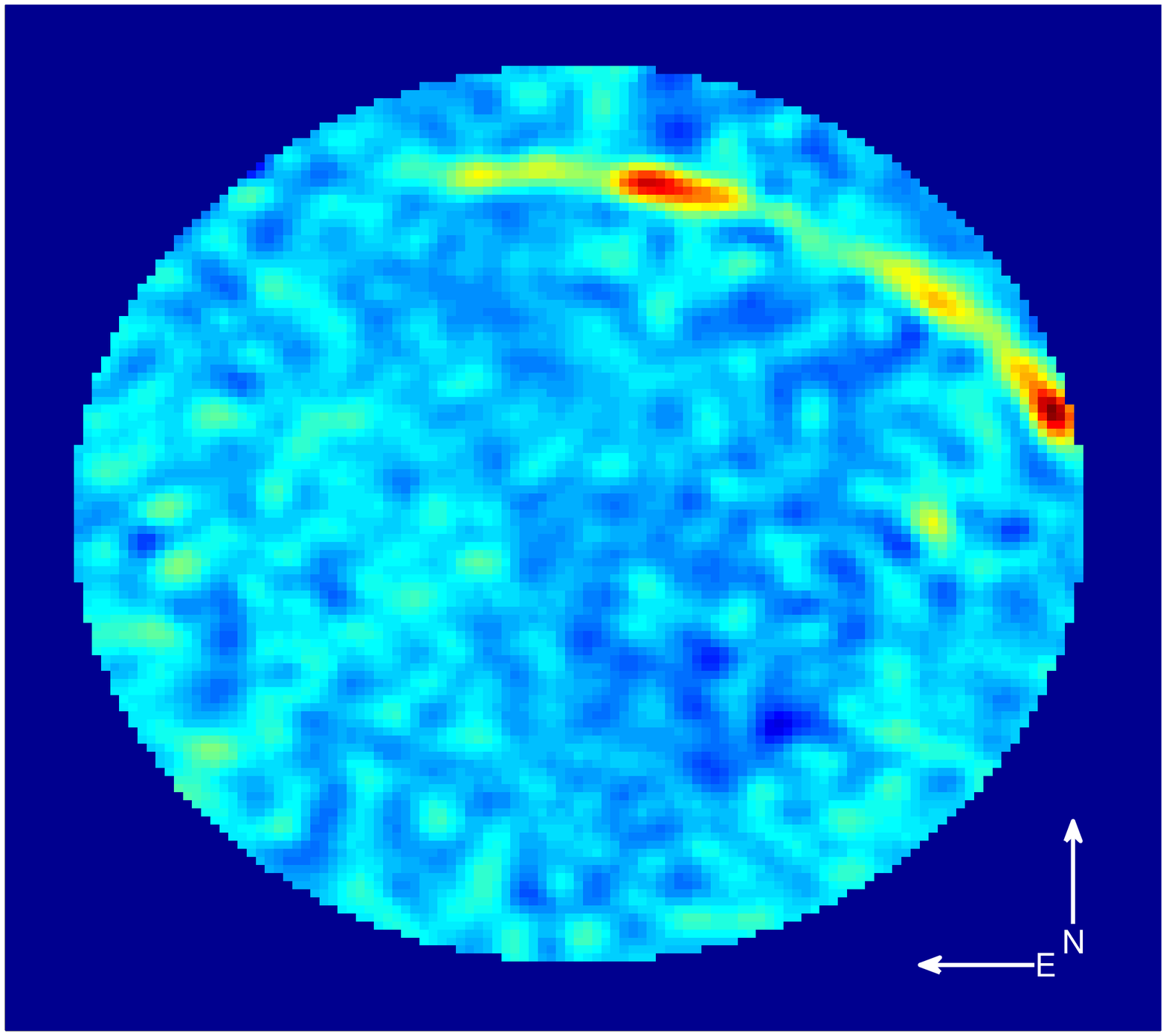}
	\caption{Image of the sky, after subtraction, of the fireball which covered 92$^{\circ}$. The edge of the circle marks a cutoff of 25$^{\circ}$ above the horizon. Strong constant sources Cassiopeia A and the Virgo A, Taurus A, and the Galactic Plane have left weak residual signals in the image. }
\end{figure}

\begin{figure}
	\centering
	\includegraphics[width = 7in]{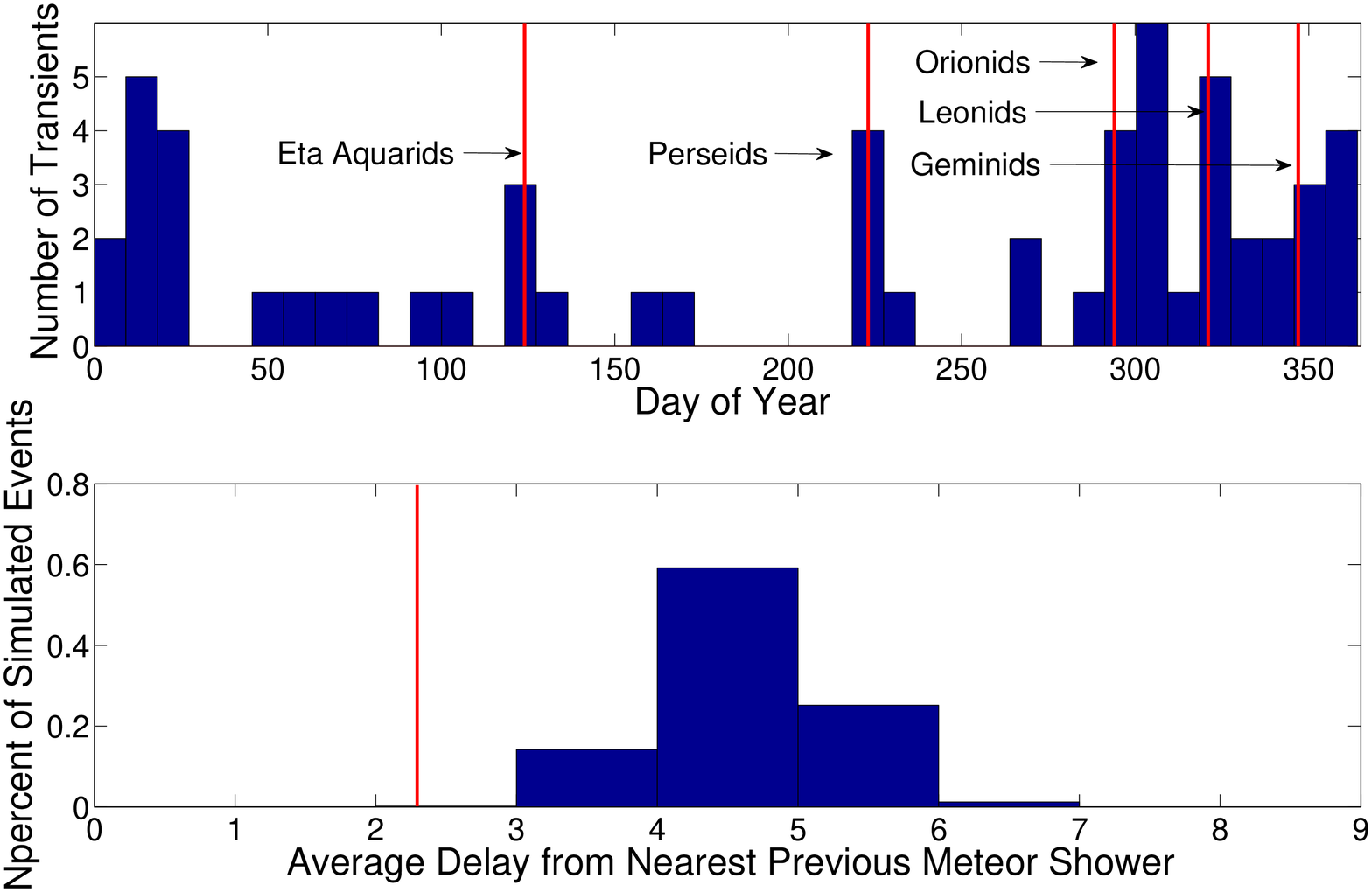}
	\caption{(Top) A histogram showing the number of events per day of year. The full 365 days of the year are grouped into 40 bins, each bin representing 9.125 days. The red lines show the dates of several major meteor showers. The large clump which occurs near January 10 - 25 does not have a major meteor shower associated with it and may be evidence of a previously unknown fireball stream. (Bottom) Shown here are the normalized results of conducting 10$^{6}$ iterations of simulated data spread over the 7028 hours of data recorded at 25.6, 37.8, and 37.9 MHz. For every iteration 45 events were generated within the actual observing times, and the mean delay of all the events to the closest meteor shower was calculated. The actual measured delay of 2.3 is shown in red.}
\end{figure}

While the two New Mexico stations are too far South East of the LWA1 to have detected the extended transient candidate, 5 of the other 44 transients correlate in both space and time to fireballs. The fireball network could not have seen the remaining 39 events because they were either too far North West or occurred during the day. Figure 3 shows a  histogram of the events, demonstrating that groups of transient detections appear around the times of meteor showers. This implies that a large fraction of the events not seen by the network are most likely meteors as well. 

\begin{figure}
	\centering
	\includegraphics[width = 7in]{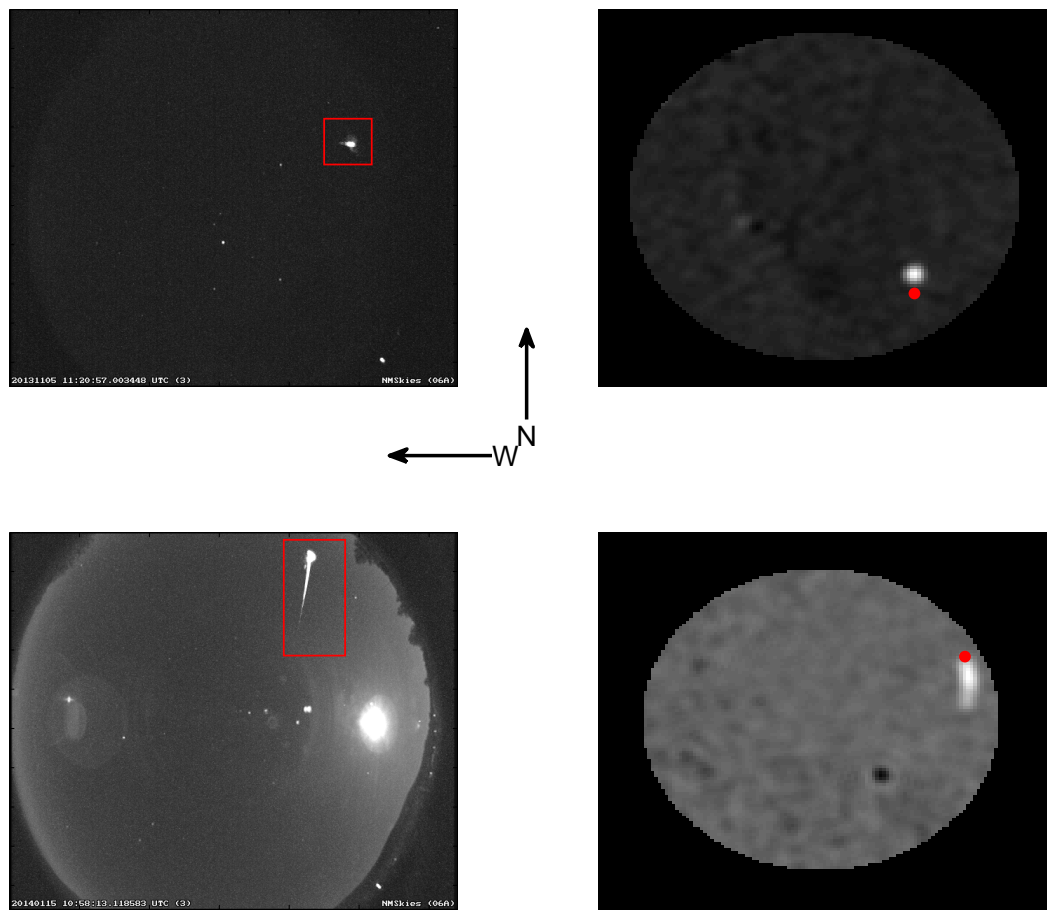}
	\caption{Comparison images from the NASA All-Sky Fireball Network station located in Mayhill, NM (Left) and PASI (Right). The red rectangles outline the fireballs in the Fireball Network images. The red dots on the PASI images show the final location of the fireball as provided by the Fireball Network. For the PASI images, the edge of the circle marks a cutoff of 25$^{\circ}$ above the horizon. East-West orientation of the PASI images have been reversed from their normal appearance to have the same orientation as the Fireball Network Images. The large bright disk in the lower left image is the moon.}
\end{figure}

The mean delay between an event and the nearest previous major meteor shower in Figure 3 is 2.3 time bins\footnote{Each time bin represents 9.125 days}. In order to quantify the probability of this distribution with respect to the meteor showers, we performed a Monte Carlo simulation. Random dates were generated and checked if they occurred during actual PASI observing time at 25.6, 37.8, or 37.9 MHz. If they did occur during times at these frequencies, they were saved into an array, otherwise they were discarded. Once 44 events were compiled, the average delay between the simulated events and the closest of the four major meteor showers was calculated. This was performed 10$^{6}$ times and a normalized histogram of the results are shown in Figure 3. The actual data lies between 2 and 3 bins which has a probability of 0.2\%.


Older events from the fireball network were used to search locations and times for our oldest data, which was taken at 25.6 MHz and was only saved as movie files which are not searchable with the image subtraction algorithm\footnote{Figure 3 only represents the 44 events found with the image subtraction algorithm, since the other 5 events were found by looking for specific fireballs rather than by a blind search. The inclusion of these would therefore skew the statistics.}. In this data an additional five extremely bright correlated events were found, one of which covered $\sim$ 65$^{\circ}$ passing through zenith and was of comparable brightness to the galactic plane. These additional events bring the total number to 49, 10 of which are correlated directly to fireballs. 

The fireballs themselves are of the more energetic variety seen by the fireball network. The velocities of all 10 were above 50 km s$^{-1}$, with an average velocity of 68 km s$^{-1}$. The observed optical emission of each event lasted for $\sim$ 1 s, and was followed shortly (usually within 1 PASI integration) by the onset of the radio emission. The four seen near 38 MHz had peak absolute visual magnitudes of $\sim$ $-$4.8, $-$6.1, $-$6.3, and $-$6.4. The six seen at 25.6 MHz had peak absolute visual magnitudes of $\sim$ $-$3.3, $-$4.1, $-$4.3, $-$4.7, $-$6.6, and $-$7.1. Meteors with visual magnitudes brighter than $-$4 occur about once every 20 hours \citep{Halliday96}. The chance overlap in space and time of 5 events\footnote{The 5 events found by using times and locations from the fireball network were not used in the calculation because they were not found randomly and therefore cannot be added to the random population of PASI transients.} found randomly in PASI data with fireballs brighter than magnitude $-$4 within the 11,000 hours of data is about 1 in 10$^{28}$. Therefore it is clear that these correlations are not mere coincidence but, indeed, connected to fireballs. Figure 4 shows images of two fireballs observed simultaneously by the Fireball Network and the LWA1.

\section{Reflection vs Emssion} The ionized trails left by meteors have long been known to scatter radio waves, and since the 1940s the use of radar echoes became a popular method for observing their trails allowing the detection of very dim as well as daytime meteors unobservable by optical instruments. A large fraction of these observations have been of specular trails, which occur when a line perpendicular to the meteor's path satisfies the reflection requirement that the angle to the transmitter equals the angle to the receiver \citep{Wislez95}. The majority of these reflections last for $<$ 1 s, have electron line densities of $\alpha < 2.4 \times 10^{12}$ cm$^{-1}$, and are known as underdense meteors. A less common variety of specular trail last for several seconds, have electron line densities of $\alpha > 2.4 \times 10^{12}$ cm$^{-1}$, and are known as overdense meteors. These are associated with larger meteors, the largest of which are associated with optical fireballs \citep{Ceplecha98,Bronshten83,McKinley61}. 

A smaller subset of meteor trail echoes are non-specular (i.e. due to scattering), which are not yet well understood. They most often occur when the radar is pointed perpendicular to the geomagnetic field, but do not necessarily satisfy the specular reflection requirement. These types of trails can last from seconds to several minutes, but are typically much weaker than specular trails and therefore high power large-aperture radars are required to detect them \citep{Bourdillon05,Sugar10,Close11}.

Meteor trail echoes are a well studied phenomenon, and their characteristics are well documented. Moreover a recent study by \citet{Helmboldt14} using 55.25 MHz analog TV broadcasting stations for meteor scatter provides a direct example of how to detect meteor echoes with the LWA1 and how to identify them in the all-sky data. In comparison, several key differences arise between the transients reported in this paper and what is expected from trail echoes. The differences are as follows:

First, typical transmitters are strongly polarized, resulting in reflections that are strongly polarized \citep{Helmboldt14,Close11,Wislez95}. However no significant amount of linear nor circular polarization has been detected from the observed transients. 

Secondly, a large portion of the RFI seen by LWA1 is narrower than the 75 kHz PASI band and is easily identifiable by its spectra \citep{Obenberger11}. Yet none of the observed transients contain any spectral features.

Thirdly, the light curves of the observed transients are consistent with each other, ranging from 30 to 150 s, showing a linear rise and a long exponential decay, and and otherwise show a smooth evolution (Figure 1). A typical reflection from an overdense trail reaches maximum brightness in just a few seconds, maintains a relatively constant average brightness while undergoing sporadic dimming and rebrightening. It then quickly decays away once it expands to the point that the density reaches the underdense criteria \citep{Ceplecha98,Wislez95,Helmboldt14}. The observed transients are also inconsistent with light curves from non-specular echoes, which vary greatly from one to the next, following no particular pattern. More importantly, however, non-specular echoes are weaker and more rare than overdense specular echoes \citep{Bourdillon05,Close11}. Therefore if the LWA1 were seeing non-specular reflections it should also see many more bright specular reflections scattering from the same transmitters.

Finally, the observed transients have azimuths and elevations consistent with the uniform distribution convolved with the LWA1 power pattern (Figure 5). This distribution implies that the sources appear in random locations with no preferable sky position. This is inconsistent with what is expected from specular echoes of man-made radio frequency interference (RFI), which should increase towards the horizon due to the increased number of incident angles with distant transmitters required for forward scattering \citep{Wislez95}. The observed pattern could be consistent with nearby transmitters. However, because the signal strength depends on the inverse-cube of the distance to the meteor, there should be many very bright nearby RFI sources on the horizon but these are not observed. This pattern is also inconsistent with non-specular reflections, which are preferentially located in a relatively small region of the sky that satisfies the requirement that the pointing vector is perpendicular to the geomagnetic field \citep{Bourdillon05,Close11}. 

\begin{figure}[!]
	\centering
	\includegraphics[width = 7in]{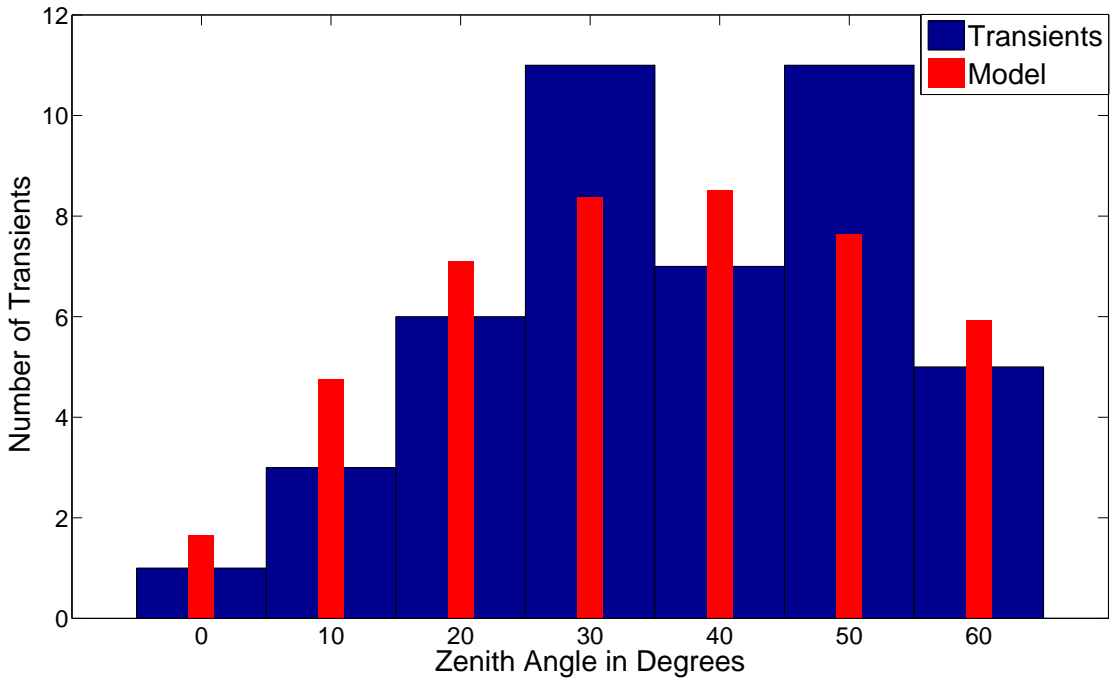}
	\caption{A histogram showing the number of events as a function of zenith angle. The blue bars show the actual measured distribution of the transients. The red bars show a model based on a uniform distribution convolved with the power pattern of the LWA1. }
\end{figure}

For these reasons it seems unlikely that forward scattering is responsible for the signals detected from fireballs. It is therefore our conclusion that fireball trails radiate at low frequencies. 

\section{Physical Constraints on the Emission}
Since PASI can only record at one center frequency at a time, there is limited spectral information for each event. Nevertheless the similarity of the light curves of fireballs recorded at 25.6, 29.9, 37.8, and 37.9 MHz implies that this emission is broad band. It also appears that they are brighter at 25.6 than at 38 MHz, suggesting a spectral slope favorable to lower frequencies. Also, there have been no detections at 52.0 or 74.0 MHz which may be the result of a sharp cutoff at higher frequencies or a steep spectrum. In either case it is clear from these observations that the emission is non-thermal, yet the exact mechanism is currently unknown. 

If a magnetic field of 10 to 15 G were present within the trail, it follows that cyclotron radiation would be emitted at the observed frequencies by the electrons in the plasma. However, the surface geomagnetic field is only 0.5 G, so this would require the generation of a strong magnetic field by a fireball, an effect that has never been observed.

Brightness temperature for a radio source can be calculated using the equation:
\begin{equation}
T_{b} = \frac{S_{\nu} \lambda^{2}}{2 k \Omega}
\end{equation}
Where $S_{\nu}$ is the flux density of the source, $\lambda$ is the observed wavelength, $k$ is the Boltzmann constant, and $\Omega$ is the solid angle of the source. The solid angle of the emitting portion of the fireball is estimated by using typical size scales reported for bright meteors trails. 

A typical fireball observed by the LWA1 has a peak flux density of 1 kJy, 20 s after first light, and most are point sources with a measured beam size of $ 4.4^{\circ}$ at 37.8 MHz. Meteor plasma trails are typically modeled as long cylinders with radii much smaller than their length. Assuming a height of 110 km and zenith angle of 40$^{\circ}$, an angular size of $ 4.4^{\circ}$ corresponds to 12 km, which is a reasonable value for the length of the emitting part of the plasma trail. However for large plasma trails a typical initial radius is 10 m, which quickly expands due to diffusion. It is estimated that at 20 s after entry, the radius of the expanding trail would be $\sim$100 m \citep{Ceplecha98,Bronshten83,McKinley61}. Given these dimensions the estimated brightness temperature is $3 \times 10^{5}$ K.  This temperature is two orders of magnitude higher than the typical peak temperature of a fireball trail, and this provides further evidence that the emission is non thermal. 

While the spectral slope has yet to be fully measured, the total energies of the radio emission are estimated by assuming a flat spectrum from 10 to 50 MHz and heights of 110 km. These assumptions yield total radio energy estimates ranging from 10$^{-4}$ to 10$^{-2}$ J, which is one part in $10^{12}$ to $10^{10}$ of the kinetic energy of a typical fireball. Further observations are in progress to better characterize the spectral properties of this emission, and to get better estimates of the total amount of energy radiated at radio frequencies.  

\section{Discussion}
Decametric radio emission from meteors has not been previously detected, but this is not the first time its existence has been discussed. \citet{Hawkins58} conducted a search for radio emission, but reported only upper limits with a 5 $\sigma$ sensitivity of $\sim 10^{8}$ Jy at 30 MHz for 1 s bursts. It is also interesting to note that in the last several decades detections of extremely low frequency (ELF\footnote{3 Hz to 3 kHz}) and very low frequency (VLF\footnote{3 kHz to 30 kHz}) emission have been reported coincident with large meteors \citep{Guha12,Keay80,Beech95}. The physical mechanism responsible for this emission is not well understood, but might be related to our detections of higher frequency emission. 

Given the vast range in energies and size scales of meteors and their corresponding plasma trails, this emission may exist at a wide range of frequencies, timescales, and energies. Investigating this emission further will yield new insights into the physics of meteors and their interaction with our atmosphere. Moreover fireballs are now a known radio transient foreground source and need to be taken into account when searching for cosmic transients. It is interesting to note that transient atmospheric phenomena, unknown to emit at radio frequencies, have been proposed as the possible source of Perytons and perhaps even Fast Radio Bursts \citep{Kulkarni14,Katz14,Burke-Spolaor11,Thornton13,Lorimer07}. In fact Katz (2014) suggests meteors as a possible source of Perytons. In this paper we reported observations of high energy meteors generating bright radio emission.

\section{Acknowledgments} Construction of the LWA1 has been supported by the Office of Naval Research under Contract N00014-07-C-0147. Support for operations and continuing development of the LWA1 is provided by the National Science Foundation under grants AST-1139963 and AST-1139974 of the University Radio Observatory program. 

This research has made use of the NASA/IPAC Extragalactic Database (NED) which is operated by the Jet Propulsion Laboratory, California Institute of Technology, under contract with the National Aeronautics and Space Administration.

Part of this research was conducted at the Jet Propulsion Laboratory, California Institute of Technology, under contract to NASA

\bibliographystyle{plainnat}

\end{document}